\documentclass[10pt]{article}
\usepackage{amsthm,amsfonts,epsfig}
\usepackage{amsmath}
\usepackage[T1]{fontenc}
\usepackage{graphicx}
\usepackage{graphics}
\usepackage{placeins}
\usepackage{here}
\usepackage{float}
\usepackage{subfigure}
\usepackage{lmodern}
\usepackage{enumerate}
\usepackage{multirow}
\usepackage{setspace}
\usepackage[a4paper, text={17cm,22.5cm},top=1cm, left=2.5cm, right=2.5cm, includeheadfoot, headheight=1cm]{geometry}
\begin{document}
  \title{Analyzing Volleyball  Data on a Compositional Regression Model Approach: An Application to the Brazilian Men's Volleyball Super League 2011/2012  Data}
	\date{}

 \author{Taciana K. O. Shimizu,\,Francisco Louzada\,and\,Adriano K. Suzuki\\
  {\small Department of Applied Mathematics \& Statistics, ICMC, University of São Paulo}\\
  {\small Av. Trabalhador Saocarlense, 400 CEP 13.566-590, São Carlos, SP, Brazil} \\
  {\small e-mail: \emph{taci\_kisaki@yahoo.com.br}, \emph{louzada@icmc.usp.br} and \emph{suzuki@icmc.usp.br}}\\
}
\maketitle

%
	\begin{abstract}
     
Volleyball has become a competitive sport with high physical and technical performance, according to the International Volleyball Federation \cite{FIVB}. Matches results are based on the players and teams'skills as technical and tactical strategies to succeed in a championship. At this point, some studies are carried out on the performance analysis of different match elements, contributing to the development of this sport. In this paper, we proposed a new approach to analyze volleyball data. The study is based on the compositional data methodology modeling in regression model. The parameters are obtained through the maximum likelihood. We performed a simulation study to evaluate the estimation procedure in compositional regression model and we illustrated the proposed methodology considering real data set of volleyball. 

\noindent
\textit{Keywords}: Compositional data, Additive log-ratio transformation, maximum likelihood estimation, volleyball League.
 \end{abstract}

\section{Introduction}
	
Volleyball is currently one of the most popular team sports worldwide. Since its origin, significant changes have been performed in the rules, contributing to the evolution of this sport. In this sense, the need for performance and skills analysis has grown considerably, not only for the players' evaluation but also for the tactical systems used by high-level teams. This type of analysis arises as an interesting research field to answer questions about practical and competition issues. 
		
In this context, studies have contributed to the development of tactical and technical strategies based on factors which allow us to determine important parameters related to matches, players and coach's decision-making. 
		
There are four different ways for a volleyball team to get score: attack, block, serve and opponent error. According to Rodriguez-Ruiz et.al \cite{RODRIGUEZ}, among these parameters, the most decisive in high-level volleyball are performance in attack, serve and block, and in the points scored with opponent error. Thus, the analysis of these performance parameters is relevant, mainly considering statistical analysis for coaching decisions.   
		
As a motivation for the present study, we consider a new approach to analyze volleyball data, and we based our study on the compositional data methodology modeling in regression model proposed by Iyengar and Dey \cite{IYENGAR}. Usually, the data (attack, block, serve and opponent error) from this type of sport has compositional restrictions, in other words, it has dependence structure. Thus, standard existing methods to analyze multivariate data under the usual assumption of multivariate normal distribution (see for example, Johnson and Wichern \cite{JOHNSON}) are not appropriate to analyze them.
		
\subsection{Literature review}		

In spite of some studies focus on the performance indicators of volleyball, they have different objectives and statistical procedures. For example, Gomez et.al \cite{GOMEZ} made a comparison between some different male team sports to assess the home advantage effect during the same period, level of competition and country (Spain). Among the sports were included baseball, basketball, handball, indoor soccer, roller hockey, rugby, soccer, volleyball and water polo. The statistical analysis consisted of applying the \textit{t} test to verify the existence of home advantage for each sport and a two-way analysis of variance to test the differences among the sports, followed by a \textit{post hoc} Tukey test to evaluate specific differences among the sports. Although the results showed significant home advantage for each sport, volleyball presented lower home advantage. 

Campos et.al \cite{CAMPOS} also investigated the advantage of playing at home and the influence of performance indicators in the game score according to the set number; however their focus was on elite women's volleyball leagues (Brazilian and Italian). Home advantage was calculated based on ratio win/loss of each team when playing at home, where score of 50\% indicated no home advantage and three dependent variables were considered, relating the final outcome to game location, the country and the number of sets of each game. The performance indicators were assessed in the points scored (home/away) and quantified the following independent variables: serve, block, attack and opponent errors. The work involved the use of Shapiro-Wilks test and variance of analysis for games with three, four and five sets according to game location (home or away), game outcome (win or lose) and league (Brazilian or Italian).
 
In the proposition of multivariate analysis, Mesquita and Sampaio \cite{MESQUITA} compared the volleyball game-related statistics by sex. They analyzed all games of the World Cup (male and female) in 2007 and then they applied descriptive procedures and a discriminant analysis to identify which variables were discriminated by sex. The variables analyzed were: serve, reception, set, attack, block and dig. 
Zirhlioglu \cite{ZIRHLIOGLU} presented a study about playing characteristics of team performance in female volleyball in Turkey. Match results data was analyzed using multidimensional scaling analysis, in which statistics of the variables were: break point, serve error, ace, reception error, excellent reception, attack error, attack block, attack points, block points. 

Sheng \cite{SHENG} presented an analysis and a prediction of the volleyball results through the data mining methods. Furthermore, data mining algorithm based on homogenous Markov model was applied.

Afonso et.al \cite{AFONSOETAL} proposed to examine predictors of the setting zone in elite-level male volleyball. The setting zone (dependent variable) was evaluated through a wide range of game actions, including serve type, server player, reception zone, serve depth, receiver player and reception type (independent variables). Multinomial logistic regression was applied to obtain the estimated likelihood of occurrence of the dependent variable. In Marcelino et.al \cite{MARCELINO} assessed the effects of quality of opposition and match status on technical and tactical volleyball indicators, i.e., block, attack, serve and set actions related to the tasks, space, players and efficacy of selected game actions. First of all, they applied cluster analysis to classify the teams into competitive levels. After this analysis, it was used the multinomial logistic regression to evaluate the association between match status and tactical indicators. 

Rodriguez-Ruiz et.al \cite{RODRIGUEZ} analyzed the terminating actions (serve, attack, block and opponent error) leading to point scoring by each team, taking them as reference indicators for observing the possibilities of winning or losing a set in a male volleyball competition. Initially, the study verified normality and homogeneity to each dataset with Kolmogorov-Smirnov test, and then they evaluated and quantified changes among study variables by groups, through the \textit{t} test for independent samples (comparison of wins and losses in sets with more than 25 points and tie break). In other cases, as analysis of total sets and 25-point sets, it was applied a non-parametric test (Mann-Whitney U test) for independent samples. Another study was also based on non-parametric tests, Patsiaouras et.al \cite{PATSIAOURAS} worked with the purpose of examining the technical elements that emerge as important factors for a match result and identifying statistically significant differences among the participant teams in the Olympic games of Beijing 2008. They used non-parametric tests (Jonckheere-Terpstra and Kolmogorov-Smirnov Z test) for the statistical analysis considering winning or losing a match as dependent variables. Through Jonckheere-Terpstra test, the results pointed out there were significant differences among the teams winning a match compared to those loosing one for the factors of service points, total reception, reception errors, and attack blocked.

The aim of this article is to consider a new approach to analyze volleyball data, in the context of the compositional data methodology modeling as regression model formulated by Iyengar and Dey \cite{IYENGAR}. Basically, we consider analyzing volleyball skills as attack, block, serves and errors of the opposite team. We will present the description of the volleyball data set in the next subsection.

\subsection{Data set}

We consider a real data set that is related to the first and second rounds matches of Brazilian Men's Volleyball Super League 2011/2012 obtained from the website \cite{CBV}. The data refers to the teams that played and won the games in these rounds, in particular the winning team points in each game which were defined as composition, and the volleyball skills as attack, block, serves and points gained from the errors of the opposite team as proportions of each composition.

The points of the winning team in each game are obtained by four components, being that we denoted $x_{i1}$ as the points proportion in attack, $x_{i2}$ the points proportion in block, $x_{i3}$ the points proportion in serve and $x_{i4}$ the points proportion in errors of the opposite team. The data set described above is shown in the Appendix (Table A).

\subsection{Paper organization}

The structure of the remaining parts of this paper is organized as follows. In Section 2, we present the definition of compositional data and the formulation of regression model applied to this type of data using Additive Log-Ratio (ALR) transformation. A simulation study of the proposed methodology is presented in Section 3. Furthermore, we present the results of the application to a real data set related to the Brazilian Men's Volleyball Super League 2011/2012. Finally, concluding remarks are given in Section 4.

\section{Regression Model with \textit{ALR} transformation}

In this section we introduce definitions about compositional data, the ALR transformation and the regression model applied in compositional data, considering the model's response variable as vector of the components proportions. 

The development of the study about compositional theory started with the contributions of Aitchison and Shen \cite{AITCHISONSHEN} and Aitchison \cite{AITCHISON1}. Since then, the applications have become greater in different areas of knowledge. Compositional data are vectors of proportions specifying $G$ fractions as a whole. Thus, for $\textbf{x} = (x_{1}, x_{2},\ldots,x_{G})'$ to be a compositional vector, we must have $x_{i}>0$, for $i = 1,\ldots,G$ and $x_{1} + x_{2} + \ldots + x_{G} = 1$. 

Compositional data often results when raw data is normalized or when data is obtained as proportions of a certain heterogeneous quantity. Standard existing methods to analyze multivariate data under the usual assumption of multivariate normal distribution (see for example, Johnson and Wichern \cite{JOHNSON}) are not appropriate to analyze compositional data, since we have compositional restrictions. 

Different modelling has been considered to analyze compositional data. The first model considered to analyze compositional data was given by the Dirichlet distribution, but this model requires that the correlation structure is wholly negative, a fact that is not observed for compositional data in which some correlations are positive (see for example, Aitchison \cite{AITCHISON1} and Aitchison \cite{AITCHISON2}).

An alternative to analyze compositional data was proposed by Aitchison \cite{AITCHISON1} considering suitable transformations from restricted sample space Simplex to well-defined real sample space. More specifically, Aitchison and Shen \cite{AITCHISONSHEN} developed the logistic-Normal class of distributions transforming the $G$ component vector $\textbf{x}$ to a vector $\textbf{y}$ in $R^{G-1}$ considering the Additive Log-Ratio (ALR) function.

According to Achcar and Obage \cite{ACHCAROBAGE}, we can consider $y_{ij}=H(x_{ij}/x_{iG})$, $i=1,...,n$ and $j=1,...,g$, being $H(\bullet)$ the chosen transformation function to assure that result vector has real components, where $x_{ij}$ represents the i-th observation for the j-th component, such that $x_{i1}>0,\ldots,x_{iG}>0$ e $\sum_{j=1}^{G}x_{ij}=1$, for $i=1,...,n$.

The ALR transformation to analyze compositional data is given by,
\begin{equation} \label{rla}
y_{ij}=H \left(\frac{x_{ij}}{x_{iG}}\right)=\mbox{log}\left(\frac{x_{ij}}{x_{iG}}\right).
\end{equation}

Assuming ALR transformation for the response variables, the regression model proposed in Iyengar and Dey \cite{IYENGAR} is given by,
\begin{equation} \label{modeloreg}
 y_{ij}=\beta_{0j}+\underline{z_{i}}\beta_{1j}+\epsilon_{ij}, 
\end{equation}

\noindent
where $y_{ij}=(y_{i1},\ldots,y_{ig})$ is a vector ($1\times g$) of response variables where $g=G-1$ and $G$ number of compositional data components; $\underline{z_{i}}$ is a vector of covariates associated to the \textit{i-th} sample; $\beta_{0j}$ is a vector ($1\times g$) intercepts; $\beta_{1j}$ is a vector ($p\times g$) regression coefficients and $\epsilon_{ij}$ are random errors, for $j=1,\ldots,G-1$ and $i=1,\ldots,n$.

The likelihood function for the parameters $\beta_{0j}, \beta_{1j}$ and $\sigma_{j}^{2}$ is given by,
\begin{eqnarray*} \label{vero}
L(\beta_{0j},\beta_{1j},\sigma_{j}|y_{ij})=\prod_{j=1}^{G-1}\left(\frac{1}{\sqrt{2\pi\sigma_{j}^{2}}}\right)^{-n/2}\mbox{exp}\left(-\frac{1}{2\sigma_{j}^{2}}\sum_{i=1}^{n}\epsilon_{ij}^{2}\right),
\end{eqnarray*}

\noindent
where
$\sum_{i=1}^{n}\epsilon_{ij}^{2}=\sum_{i=1}^{n}(y_{ij}-\beta_{0j}-\beta_{1j}\underline{z_{i}})^{2}$, for $j=1,\ldots,G-1$ and $i=1,\ldots,n$.

Following Migon et.al \cite{MIGON}, the approximate $(1-\alpha)$ 100\% confidence intervals for the parameters $\beta_{0j}, \beta_{1j}, \sigma_{j}$ are given by,
\begin{eqnarray*}
\widehat{\beta}_{0j} &\pm& \xi_{\delta/2}\sqrt{Var(\widehat{\beta}_{0j})}, \\
\widehat{\beta}_{1j} &\pm& \xi_{\delta/2}\sqrt{Var(\widehat{\beta}_{1j})} \quad \mbox{and}\\
\widehat{\sigma}_{j} &\pm& \xi_{\delta/2}\sqrt{Var(\widehat{\sigma}_{j})}, \\
\end{eqnarray*}

\noindent
where $\xi_{\delta/2}$ is the upper $\delta/2$ percentile of standard Normal distribution and $j=1,\ldots,3$.

\section{Application}

In this section, we present a simulation study in compositional data and the results from the analysis of a real data set to illustrate an application of the proposed methodology, in particular, data related to proportions of the points volleyball teams. 

\subsection{Simulation study}

The simulation study was considered to illustrate the proposed methodology when we have unbalanced compositional data to four components. The simulated data was generated randomly from $U(0,1)$ according to unbalanced proportions (components: 0.50, 0.15, 0.05, 0.30). The covariate associated in the model (\ref{modeloreg}) is defined by,
\begin{eqnarray} \label{covariate}
 z_{i}=\left \{
\begin{array}{ll}
  1,& \mbox{if the player who scored in the i-th game is of the win team} \\
  0,& \mbox{otherwise}.  
\end{array}\right.
\end{eqnarray}

\noindent
and it was generated through $z \sim$ Bernoulli $(0.5)$. 

Moreover, the simulation study was based on 1000 samples generated for each case mentioned above. It was chosen sample sizes $n=70, 100$ and $150$ (number of volleyball games). 
The parameters values were fixed as follows: $\beta_{01}=0.5,\beta_{02}=-0.62,\beta_{03}=-1.68$, $\beta_{11},\beta_{12},\beta_{13}=-0.05$ and $\sigma_{1}=0.31,\sigma_{2}=0.41,\sigma_{3}=0.75$.

The software R was used to perform the simulations. Table \ref{tabsim2} presents the simulation results, such as mean, bias, mean square error (MSE) and coverage probability (CP). 

The simulations results showed that the MSE and bias decreases with the sample size, while the CP is stable, near the nominal coverage.

\begin{table}[!h] 
\centering{\caption{Simulated data. Mean, bias, MSE and CP for the maximum likelihood estimates based on 1.000 generated samples.}  
\footnotesize
\vspace*{0.2cm}
\begin{tabular}{c|cccccc}
\hline	
\multirow{2}{*}{Data} & Sample & \multirow{2}{*}{Parameter} & \multirow{2}{*}{Mean} &	\multirow{2}{*}{Bias}  &	\multirow{2}{*}{MSE} & \multirow{2}{*}{CP} \\
 & Size & & & & & \\
\hline
&		&	$\beta_{01}$	&	0.55937	&	0.05937	&	0.00484	&	0.999	\\
	&		&	$\beta_{02}$	&	-0.66107	&	-0.04107	&	0.00482	&	0.996	\\
	&		&	$\beta_{03}$	&	-1.71249	&	-0.03249	&	0.01519	&	0.845	\\
	&		&	$\beta_{11}$	&	-0.00381	&	0.04619	&	0.00482	&	0.997	\\
	&	$n=70$	&	$\beta_{12}$	&	-0.00499	&	0.04501	&	0.00838	&	0.953	\\
	&		&	$\beta_{13}$	&	-0.00600	&	0.04400	&	0.02937	&	0.697	\\
	&		&	$\sigma_{1}$	&	0.20948	&	-0.10052	&	0.01166	&	0.992	\\
	&		&	$\sigma_{2}$	&	0.32658	&	-0.08342	&	0.00786	&	0.999	\\
	&		&	$\sigma_{3}$	&	0.70435	&	-0.04565	&	0.00475	&	0.987	\\ \cline{2-7}
	&		&	$\beta_{01}$	&	0.55960	&	0.05960	&	0.00439	&	0.988	\\
	&		&	$\beta_{02}$	&	-0.66270	&	-0.04270	&	0.00385	&	0.990	\\
	&		&	$\beta_{03}$	&	-1.70867	&	-0.02867	&	0.01055	&	0.847	\\
	&		&	$\beta_{11}$	&	-0.00211	&	0.04789	&	0.00422	&	0.989	\\
Unbalanced data &	$n=100$	&	$\beta_{12}$ 	&	-0.00185	&	0.04815	&	0.00685	&	0.929	\\
 	&		&	$\beta_{13}$	&	-0.00854	&	0.04146	&	0.02198	&	0.660	\\
	&		&	$\sigma_{1}$	&	0.21027	&	-0.09973	&	0.01191	&	0.988	\\
	&		&	$\sigma_{2}$	&	0.32972	&	-0.08027	&	0.00677	&	0.999	\\
	&		&	$\sigma_{3}$	&	0.70926	&	-0.04074	&	0.00365	&	0.992	\\ \cline{2-7}
	&		&	$\beta_{01}$	&	0.55753	&	0.05753	&	0.00392	&	0.984	\\
	&		&	$\beta_{02}$	&	-0.66294	&	-0.04294	&	0.00324	&	0.985	\\
	&		&	$\beta_{03}$	&	-1.71676	&	-0.03676	&	0.00785	&	0.828	\\
	&		&	$\beta_{11}$	&	0.00118	&	0.05118	&	0.00394	&	0.971	\\
	&	$n=150$	&	$\beta_{12}$	&	-0.00009	&	0.04991	&	0.00563	&	0.895	\\
	&		&	$\beta_{13}$	&	0.00359	&	0.05359	&	0.01727	&	0.624	\\
	&		&	$\sigma_{1}$	&	0.21458	&	-0.09542	&	0.00924	&	0.967	\\
	&		&	$\sigma_{2}$	&	0.33079	&	-0.07920	&	0.00651	&	0.992	\\
	&		&	$\sigma_{3}$	&	0.71310	&	-0.03690	&	0.00275	&	0.973	\\
\hline
\end{tabular}
\label{tabsim2}}
\end{table}

\newpage
\subsection{Real data application}

We consider the real data set was presented in the Section 1.2 and we apply the compositional data methodology in this data set considering as components proportions the win team points in 128 games of Brazilian Men's Volleyball Super League 2011/2012. This study based on the four components: attack ($x_{i1}$), block ($x_{i2}$), serve ($x_{i3}$) and errors of the opposite team ($x_{i4}$), for $i=1,\ldots,128$.
  
First of all, we consider the ALR transformation in the response variable vector $\underline{y_{i}}$ presents in (\ref{rla}). Thus, we have that,
\begin{eqnarray*}
y_{i1}=\mbox{log}\left(\frac{x_{i1}}{x_{i4}}\right), \quad y_{i2}=\mbox{log}\left(\frac{x_{i2}}{x_{i4}}\right) \quad \mbox{and} \quad y_{i3}=\mbox{log}\left(\frac{x_{i3}}{x_{i4}}\right).
\end{eqnarray*}

According to the model presented in (\ref{modeloreg}), the regression model obtained to transformed data $y_{i1}$, $y_{i2}$ e $y_{i3}$, is given by,
\begin{eqnarray} \label{model4.1}
y_{i1}&=&\beta_{01}+\beta_{11}z_{i}+\epsilon_{i1}, \nonumber\\
y_{i2}&=&\beta_{02}+\beta_{12}z_{i}+\epsilon_{i2} \quad \mbox{and} \\
y_{i3}&=&\beta_{03}+\beta_{13}z_{i}+\epsilon_{i3}, \nonumber
\end{eqnarray}

\noindent
where the covariate associated with \textit{i-th} game is given by (\ref{covariate}), $y_{ij}$ represents the transformed proportion of \textit{j-th} component (attack, block, serve and errors of the opposite team) in the \textit{i-th} game, $\beta_{0j}$ represents the mean of the points proportion in the \textit{j-th} component related with the component $x_{i4}$ (errors of the opposite team) for the team did not win the Super League, $\beta_{1j}$ indicates if there is or not associated covariate effect to the \textit{i-th} game and $\epsilon_{ij}$ represents the errors vector assumed to be independent random variables with a normal distribution $N(0,\sigma_{j}^{2})$.

Furthermore, we can estimate the true components proportions $\alpha_{i1},\alpha_{i2},\alpha_{i3},\alpha_{i4}$, where $\alpha_{i1}+\alpha_{i2}+\alpha_{i3}+\alpha_{i4}=1$, with $\alpha_{i1}>0,\alpha_{i2}>0,\alpha_{i3}>0$ e $\alpha_{i4}>0$, when we used the ALR transformation, thus we have the following relations based in (\ref{model4.1}),
\begin{eqnarray} \label{log4.1}
\mbox{log} \left(\frac{\alpha_{i1}}{\alpha_{i4}}\right)&=&\beta_{01}+\beta_{11}z_{i}, \nonumber \\
\mbox{log} \left(\frac{\alpha_{i2}}{\alpha_{i4}}\right)&=&\beta_{02}+\beta_{12}z_{i} \quad \mbox{and} \\
\mbox{log} \left(\frac{\alpha_{i3}}{\alpha_{i4}}\right)&=&\beta_{03}+\beta_{13}z_{i}. \nonumber
\end{eqnarray}

After some calculation, the estimate to obtain the true components proportions through ALR transformation is given by,
\begin{eqnarray} \label{propcomp1}
\alpha_{ij}&=&\frac{e^{\beta_{0j}+\beta_{1j}z_{i}}}{\left(1+e^{\beta_{01}+\beta_{11}z_{i}}+e^{\beta_{02}+\beta_{12}z_{i}}+e^{\beta_{03}+\beta_{13}z_{i}}\right)} \quad \mbox{and}\\
\alpha_{i4}&=&\frac{1}{\left(1+e^{\beta_{01}+\beta_{11}z_{i}}+e^{\beta_{02}+\beta_{12}z_{i}}+e^{\beta_{03}+\beta_{13}z_{i}}\right)}, \nonumber
\end{eqnarray}

\noindent
where $i=1,\ldots,128$ e $j=1,2,3$. According Achcar and Obage \cite{ACHCAROBAGE}, the parametration $\alpha_{i1}>0$, $\alpha_{i2}>0$, $\alpha_{i3}>0$ e $\alpha_{i4}>0$ can be used to obtain inferences about the compositions in each covariate value.

The joint probability density function for the parameters $\underline{\nu}=(\beta_{01},\beta_{02},\beta_{03}$, $\beta_{11},\beta_{12},\beta_{13},\sigma_{1}^{2},\sigma_{2}^{2},\sigma_{3}^{2})$ can be expressed by,
\begin{equation}
f(y_{i1},y_{i2},y_{i3}|\underline{\nu})=\prod_{j=1}^{3}\frac{1}{\sqrt{2\pi\sigma_{j}^{2}}}\mbox{exp}\left[-\frac{1}{2\sigma_{j}^{2}}(y_{ij}-\beta_{0j}-\beta_{1j}z_{i})^{2}\right],
\end{equation} 

\noindent
such that, $y_{ij}\sim\mbox{N}(\beta_{0j}+\beta_{1j}z_{i},\sigma_{j}^{2})$, for $j=1,2,3$ and $i=1,\ldots,128$.

Thus, assuming the model (\ref{model4.1}), the likelihood function for the parameters $\underline{\nu}=(\beta_{01},\beta_{02},\beta_{03}$, $\beta_{11},\beta_{12},\beta_{13},\sigma_{1}^{2},\sigma_{2}^{2},\sigma_{3}^{2})$ is given by,
\begin{eqnarray*} \label{vero4.1}
L(\underline{\nu}|y_{i1},y_{i2},y_{i3})=\prod_{j=1}^{3}\left(\frac{1}{\sqrt{2\pi\sigma_{j}^{2}}}\right)^{-n/2}\mbox{exp}\left(-\frac{1}{2\sigma_{j}^{2}}\sum_{i=1}^{n}\epsilon_{ij}^{2}\right),
\end{eqnarray*}

\noindent
where
$\sum_{i=1}^{128}\epsilon_{ij}^{2}=\sum_{i=1}^{128}(y_{ij}-\beta_{0j}-\beta_{1j}z_{i})^{2}$, for $j=1,2,3$ and $i=1,\ldots,128$.

To analyze the volleyball data, we used the software R to estimate the model parameters (\ref{model4.1}). Table \ref{tabclass1} presents the results of the model fitting to volleyball data.

\begin{table}[!ht] 
\centering\caption{Results for the compositional regression model with ALR transformation.}  
\vspace*{0.2cm}
\begin{tabular}{cccl}
\hline	
\multirow{2}{*}{Parameter} & \multirow{2}{*}{Estimate} &	Standard  &	Confidence  \\
  & & Deviation & Interval (95\%) \\
\hline
$\beta_{01}$	&	0.468	&	0.038	&	(0.394; 0.542)	\\
$\beta_{02}$	&	-1.168	&	0.066	&	(-1.296; -1.039)	\\
$\beta_{03}$	&	-2.072	&	0.087	&	(-2.244; -1.901)	\\
$\beta_{11}$	&	0.196	&	0.046	&	(0.105; 0.286)	\\
$\beta_{12}$	&	0.141	&	0.080	&	(-0.016; 0.298)	\\
$\beta_{13}$	&	0.262	&	0.106	&	(0.053; 0.470)	\\
$\sigma_{1}$	&	0.245	&	0.015	&	(0.215; 0.275)	\\
$\sigma_{2}$	&	0.426	&	0.027	&	(0.374; 0.478)	\\
$\sigma_{3}$	&	0.566	&	0.035	&	(0.496; 0.635)	\\
\hline
\end{tabular}
\label{tabclass1}
\end{table}

The results indicate there was significant effect of the covariate associated with the proportions attack and serve points ($\beta_{12}$ and $\beta_{13}$). In other words, the player who scored in the game belongs to the winning team and helps it in the following skills: attack and serve points. 

Although we understand that our multivariate approach, directed by a method for compositional data, it is the most adequate one for this kind of data since its nature is multivariate, just for the sake of comparison, we also consider a usual univariate analysis, by analyzing each component separately. Table \ref{tabprop1} presents the results on the estimated  proportions for the components attack, block, serve and errors of the opposite team according to the covariate values ($z=0$ or $z=1$), considering an ALR transformation.  

\begin{table}[!ht] 
\centering\caption{Proportions estimates for the components.}  
\vspace*{0.2cm}
\begin{tabular}{c|cc||cc}
\hline	
\multirow{2}{*}{Proportions} &	\multirow{2}{*}{$z=0$} & Confidence & \multirow{2}{*}{$z=1$} & Confidence \\
 & & Interval (95\%) & & Interval (95\%) \\
\hline
$\alpha_{1}$	&	0.526	&	(0.518; 0.533)	&	0.561 	&	(0.544; 0.571)	\\
$\alpha_{2}$	&	0.102	&	(0.096; 0.110)	&	0.103 	&	(0.089; 0.119)	\\
$\alpha_{3}$	&	0.041	&	(0.037; 0.046)	&	0.047 	&	(0.037; 0.060)	\\
$\alpha_{4}$	&	0.330	&	(0.310; 0.349)	&	0.289 	&	(0.250; 0.330)	\\
\hline
\end{tabular}
\label{tabprop1}
\end{table}

We observe the confidence intervals overlap. This is evidence that the estimated proportions for the components (attack, block and serve) are not statistically different in regard to the player who scored belonging or not to the winning team. This result contrasts with the outcome that was obtained when we analyzed this data set considering its natural multivariate structure.

Thus, from the practical point of view, the results indicated that the impact of our new approach is relevant, while analyzing each component separately leads to erroneous results since they do not take into account the structure of correlation between components.

\section{Concluding remarks}

We presented a new approach to the volleyball data, in which our study was based on the compositional data methodology modeling in regression model.

An important contribution of this study was to point out that when the player who scored more in the match belongs to the winner team, he helps with specific skills (attack, block, serve and errors of the opposite team) individually. However, a multivariate data structure must be considered. The interesting parameters of the model were estimated through the maximum likelihood method, where we observed significant effect of the covariate associated with the proportions of attack and serve points. 

These results are corroborated by the conclusion presented in Silva et.al \cite{SILVA}. In this study, it was suggested that competition must be evaluated in terms of performance details and this information needs to be used as strategies that the teams need in order to succeed in the match (victory). 
 
Furthermore, we carried out a simulation study to compositional data, considering bias, MSE and CP for different samples sizes, and discovered that the MSE and bias decreases with the sample size, while the CP is stable, near the nominal coverage.

In the real example we observed the contrast between the results obtained by considering our multivariate approach and the usual univariate one,  pointing out the advantage of considering the natural multivariate structure of the data.

\newpage
\section*{Appendix}

\subsection*{Table A. Matches of Brazilian Men's Volleyball Super League 2011/2012.}

\begin{table}[!h]
\scriptsize
\centering
\vspace*{0.1cm}
\begin{tabular}{c|ccccc||c|ccccc}
\hline	 
Matches	&	\% attack	&	\% block	&	\% serve	&	\% errors	&	z &	Matches	&	\% attack	&	\% block	&	\% serve	&	\% errors	&	z \\
\hline
1	&	48.00	&	12.00	&	2.67	&	37.33	&	1	&	65	&	53.27	&	9.35	&	5.61	&	31.78	&	1	\\
2	&	53.06	&	14.29	&	7.14	&	25.51	&	1	&	66	&	51.09	&	6.52	&	4.35	&	38.04	&	1	\\
3	&	44.00	&	13.33	&	8.00	&	34.67	&	0	&	67	&	55.26	&	6.58	&	6.58	&	31.58	&	1	\\
4	&	52.63	&	14.74	&	7.37	&	25.26	&	0	&	68	&	55.10	&	12.24	&	3.06	&	29.59	&	0	\\
5	&	56.00	&	8.00	&	5.33	&	30.67	&	1	&	69	&	64.94	&	6.49	&	5.19	&	23.38	&	1	\\
6	&	65.63	&	10.16	&	2.34	&	21.88	&	0	&	70	&	57.33	&	10.67	&	10.67	&	21.33	&	1	\\
7	&	54.67	&	14.67	&	2.67	&	28.00	&	1	&	71	&	56.70	&	12.37	&	7.22	&	23.71	&	0	\\
8	&	50.00	&	12.50	&	9.62	&	27.88	&	1	&	72	&	44.74	&	9.21	&	7.89	&	38.16	&	0	\\
9	&	52.58	&	15.46	&	3.09	&	28.87	&	0	&	73	&	62.50	&	13.54	&	4.17	&	19.79	&	1	\\
10	&	57.33	&	13.33	&	4.00	&	25.33	&	1	&	74	&	56.52	&	5.22	&	6.09	&	32.17	&	1	\\
11	&	56.60	&	11.32	&	6.60	&	25.47	&	0	&	75	&	53.19	&	14.89	&	4.26	&	27.66	&	0	\\
12	&	49.33	&	9.33	&	14.67	&	26.67	&	1	&	76	&	57.33	&	9.33	&	6.67	&	26.67	&	1	\\
13	&	56.67	&	8.33	&	6.67	&	28.33	&	0	&	77	&	50.67	&	18.67	&	2.67	&	28.00	&	1	\\
14	&	66.67	&	4.00	&	1.33	&	28.00	&	1	&	78	&	53.95	&	10.53	&	6.58	&	28.95	&	0	\\
15	&	56.70	&	10.31	&	3.09	&	29.90	&	0	&	79	&	44.00	&	16.00	&	2.67	&	37.33	&	1	\\
16	&	62.67	&	6.67	&	4.00	&	26.67	&	1	&	80	&	48.00	&	5.33	&	6.67	&	40.00	&	1	\\
17	&	49.35	&	5.19	&	3.90	&	41.56	&	1	&	81	&	56.00	&	8.00	&	2.67	&	33.33	&	0	\\
18	&	48.00	&	13.33	&	6.67	&	32.00	&	1	&	82	&	57.84	&	11.76	&	2.94	&	27.45	&	0	\\
19	&	56.12	&	14.29	&	4.08	&	25.51	&	0	&	83	&	47.25	&	12.09	&	4.40	&	36.26	&	1	\\
20	&	44.00	&	9.33	&	5.33	&	41.33	&	0	&	84	&	59.81	&	11.21	&	3.74	&	25.23	&	0	\\
21	&	54.95	&	10.81	&	5.41	&	28.83	&	1	&	85	&	53.06	&	11.22	&	5.10	&	30.61	&	1	\\
22	&	59.34	&	15.38	&	3.30	&	21.98	&	0	&	86	&	54.55	&	7.27	&	4.55	&	33.64	&	0	\\
23	&	51.52	&	12.12	&	1.01	&	35.35	&	0	&	87	&	59.18	&	6.12	&	3.06	&	31.63	&	1	\\
24	&	57.69	&	14.10	&	5.13	&	23.08	&	0	&	88	&	46.88	&	13.54	&	3.13	&	36.46	&	1	\\
25	&	53.33	&	14.67	&	2.67	&	29.33	&	0	&	89	&	57.50	&	11.67	&	1.67	&	29.17	&	0	\\
26	&	55.45	&	13.86	&	5.94	&	24.75	&	1	&	90	&	48.00	&	20.00	&	8.00	&	24.00	&	0	\\
27	&	48.04	&	10.78	&	3.92	&	37.25	&	0	&	91	&	54.00	&	13.00	&	2.00	&	31.00	&	1	\\
28	&	48.00	&	17.33	&	9.33	&	25.33	&	0	&	92	&	54.67	&	6.67	&	8.00	&	30.67	&	1	\\
29	&	57.33	&	8.00	&	6.67	&	28.00	&	1	&	93	&	56.12	&	7.14	&	3.06	&	33.67	&	1	\\
30	&	53.78	&	8.40	&	5.04	&	32.77	&	1	&	94	&	56.00	&	8.00	&	8.00	&	28.00	&	1	\\
31	&	44.74	&	10.53	&	7.02	&	37.72	&	1	&	95	&	50.67	&	13.33	&	6.67	&	29.33	&	0	\\
32	&	53.57	&	13.10	&	3.57	&	29.76	&	1	&	96	&	52.00	&	12.00	&	1.33	&	34.67	&	1	\\
33	&	59.41	&	5.94	&	6.93	&	27.72	&	1	&	97	&	58.06	&	10.75	&	2.15	&	29.03	&	0	\\
34	&	59.21	&	9.21	&	6.58	&	25.00	&	0	&	98	&	45.33	&	6.67	&	2.67	&	45.33	&	0	\\
35	&	51.04	&	13.54	&	6.25	&	29.17	&	0	&	99	&	51.90	&	15.19	&	2.53	&	30.38	&	0	\\
36	&	50.51	&	13.13	&	3.03	&	33.33	&	0	&	100	&	52.78	&	15.74	&	1.85	&	29.63	&	1	\\
37	&	55.34	&	16.50	&	4.85	&	23.30	&	0	&	101	&	54.29	&	9.52	&	7.62	&	28.57	&	0	\\
38	&	61.33	&	4.00	&	2.67	&	32.00	&	0	&	102	&	50.67	&	14.67	&	5.33	&	29.33	&	0	\\
39	&	48.98	&	17.35	&	2.04	&	31.63	&	1	&	103	&	48.00	&	20.00	&	2.67	&	29.33	&	1	\\
40	&	55.32	&	9.57	&	5.32	&	29.79	&	1	&	104	&	62.03	&	6.33	&	2.53	&	29.11	&	1	\\
41	&	56.25	&	7.29	&	5.21	&	31.25	&	1	&	105	&	40.00	&	12.00	&	9.33	&	38.67	&	0	\\
42	&	56.19	&	7.62	&	2.86	&	33.33	&	0	&	106	&	59.09	&	8.18	&	4.55	&	28.18	&	1	\\
43	&	51.72	&	8.62	&	6.03	&	33.62	&	1	&	107	&	49.33	&	9.33	&	8.00	&	33.33	&	1	\\
44	&	49.46	&	13.98	&	3.23	&	33.33	&	1	&	108	&	58.25	&	6.80	&	3.88	&	31.07	&	0	\\
45	&	47.27	&	11.82	&	5.45	&	35.45	&	0	&	109	&	60.78	&	9.80	&	6.86	&	22.55	&	0	\\
46	&	57.33	&	13.33	&	1.33	&	28.00	&	0	&	110	&	55.77	&	13.46	&	1.92	&	28.85	&	0	\\
47	&	56.84	&	14.74	&	3.16	&	25.26	&	0	&	111	&	56.00	&	8.00	&	4.00	&	32.00	&	0	\\
48	&	60.61	&	10.10	&	3.03	&	26.26	&	0	&	112	&	55.88	&	10.78	&	5.88	&	27.45	&	0	\\
49	&	60.18	&	10.62	&	3.54	&	25.66	&	0	&	113	&	64.13	&	5.43	&	7.61	&	22.83	&	0	\\
50	&	52.43	&	9.71	&	3.88	&	33.98	&	0	&	114	&	46.67	&	8.00	&	9.33	&	36.00	&	1	\\
51	&	50.67	&	13.33	&	4.00	&	32.00	&	0	&	115	&	56.14	&	8.77	&	4.39	&	30.70	&	1	\\
52	&	63.30	&	8.26	&	4.59	&	23.85	&	1	&	116	&	49.00	&	9.00	&	6.00	&	36.00	&	1	\\
53	&	54.46	&	4.95	&	2.97	&	37.62	&	1	&	117	&	54.67	&	10.67	&	5.33	&	29.33	&	1	\\
54	&	56.25	&	8.33	&	4.17	&	31.25	&	0	&	118	&	48.65	&	18.02	&	4.50	&	28.83	&	1	\\
55	&	69.89	&	5.38	&	5.38	&	19.35	&	0	&	119	&	64.22	&	7.34	&	1.83	&	26.61	&	1	\\
56	&	65.82	&	15.19	&	3.80	&	15.19	&	0	&	120	&	56.58	&	7.89	&	11.84	&	23.68	&	1	\\
57	&	57.89	&	5.26	&	11.84	&	25.00	&	1	&	121	&	56.38	&	14.89	&	5.32	&	23.40	&	0	\\
58	&	36.84	&	12.63	&	7.37	&	43.16	&	0	&	122	&	48.45	&	15.46	&	5.15	&	30.93	&	0	\\
59	&	50.00	&	14.42	&	1.92	&	33.65	&	0	&	123	&	52.83	&	6.60	&	6.60	&	33.96	&	0	\\
60	&	52.08	&	6.25	&	5.21	&	36.46	&	0	&	124	&	60.00	&	5.33	&	9.33	&	25.33	&	0	\\
61	&	53.33	&	13.33	&	5.33	&	28.00	&	0	&	125	&	59.63	&	9.17	&	1.83	&	29.36	&	1	\\
62	&	44.00	&	16.00	&	10.67	&	29.33	&	1	&	126	&	54.67	&	10.67	&	1.33	&	33.33	&	1	\\
63	&	58.67	&	10.67	&	8.00	&	22.67	&	1	&	127	&	54.67	&	6.67	&	6.67	&	32.00	&	0	\\
64	&	54.00	&	8.00	&	5.00	&	33.00	&	0	&	128	&	46.91	&	6.17	&	4.94	&	41.98	&	1	\\
\hline
\end{tabular}
\label{data}
\end{table}

\end{document}